\documentclass[slac_one]{revtex4}
\usepackage{graphicx}
\usepackage{fancyhdr}
\begin{document}

\title{DMTPC: a new apparatus for  directional detection of Dark Matter } 

%

%
\author{G.~Sciolla\footnote{Corresponding author}, A.~Lee, J.~Battat, T.~Caldwell, B.~Cornell, D.~Dujmic, P.~Fisher, S.~Henderson, R.~Lanza,  J.~Lopez, A.~Kaboth, G.~Kohse,  J.~Monroe, T.~Sahin, R.~Vanderspek, R.~Yamamoto,  H.~Yegoryan} 
\affiliation{Massachusetts Institute of Technology, Cambridge, MA 02421, USA}

\author{S.~Alhen, D.~Avery, K.~Otis, A.~Roccaro,   H.~Tomita} 
\affiliation{Boston University, Boston, MA 02215, USA}

\author{A.~Dushkin , H.~Wellenstein} 
\affiliation{Brandeis University, Waltham, MA 02454, USA}

\begin{abstract}
Directional detection of Dark Matter  allows for unambiguous 
direct detection of WIMPs as well as 
discrimination between various Dark Matter models in our galaxy. 
The DMTPC detector is a low-pressure TPC with optical readout 
designed for directional direct detection of WIMPs. 
By using $CF_4$ gas as the active material, 
the detector also has excellent 
sensitivity to spin-dependent 
interactions of Dark Matter on protons. 

\end{abstract}

\maketitle

\thispagestyle{fancy}

\section{Directional Dark Matter detection} 
Directional detection allows for unambiguous observation of Dark Matter (DM) even in presence of insidious backgrounds. 
When a Weakly Interacting Massive Particle (WIMP) collides with a nucleus in the active mass of the detector, the direction of the nuclear recoil 
encodes the direction of the incident particle. For detectors consisting of a low-pressure gas (about 50 torr), the typical length of such a recoil 
is 1-2 mm, which is sufficiently long to be reconstructed. 
The simplest models of the distribution of WIMPs in our Galaxy suggest that the orbital motion of the Sun about the Galactic center will cause an 
Earth-bound observer to experience a WIMP wind with speed 220 km/s (the galacto-centric velocity of the Sun), originating from the direction of the Sun's motion.
Because the direction of this wind is inclined by 42$^o$ with respect to the rotational axis of the Earth, the average DM direction changes by almost 90$^o$ every 12 hours~\cite{Ref1}. An ability to measure such a direction would allow for a powerful suppression of insidious backgrounds (e.g. neutrons) as well as a unique instrument to test local DM halo models.

\section{Detector Concept}
Our detector consists of a low-pressure TPC with optical readout. The target gas is $CF_4$, whose spin 1/2 fluorine nuclei provide the ideal target material 
to detect spin-dependent interactions~\cite{Ref8}. In addition,  $CF_4$ has high scintillation efficiency and low diffusion, and is ideal for underground 
detectors because it is non-toxic and non-flammable. 

When an incoming WIMP collides with a  $CF_4$ molecule at the design pressure of 50 torr, the emitted fluorine nucleus recoils 1-2 mm, ionizing the surrounding gas. The primary electrons drift toward the amplification region, where charge multiplication takes place. To achieve 2D resolution, the amplification plane is built out of woven meshes~\cite{Ref2}. In the avalanche, scintillation photons are produced together with the electrons in the ratio of 1:3~\cite{Ref3}. 
Such photons are imaged by a CCD camera triggered by a PMT or electronic readout of the charge collected on the meshes. 

The DMTPC detector is designed to measure the following quantities:
\begin{itemize}
\item{} total light collected by the CCD, which is proportional to 
the energy of the nuclear recoil; 
\item{}length of the recoil track projected onto the amplification plane; 
\item{} length of the recoil track perpendicular to the amplification plane inferred from the width of the PMT signal;
\item versus of the nuclear recoil (``head-tail'') as determined by the shape of the energy loss along the recoil track (dE/dx).  At the typical energy of these recoils, dE/dx decreases as the track slows down.
\end{itemize} 
The simultaneous measurement of energy and length of the recoil effectively rejects electrons and $\alpha$ tracks.

\section{Prototypes and detector performance}
Several prototypes  (figure 1, top left)  have been built to prove the DMTPC detector concept. Alpha tracks from $^{241}$Am and low-energy neutrons from $^{252}$Cf are used to calibrate the device and measure its performance ~\cite{Ref4}. 
For nuclear recoils of 100 keV, where the angular resolution is  15$^o$, we have achieved an energy resolution of   $\approx$ 10\%. 
Typical gas gains are $\approx$ 10$^4$-10$^5$. The intrinsic spatial resolution is  of the order of 100 $\mu$m, adequate to image recoils of 1-2 mm length with typical diffusion of 200-700 $\mu$m.
The detector performance has been simulated using a combination of SRIM~\cite{Ref5}, CASINO~\cite{Ref6}, and GEANT4 ~\cite{Ref7}. The data-MC agreement is better than 10\%.

A 10-liter  detector~\cite{denis} has been built and is being commissioned in the laboratory. 
Underground operation is expected in early 2009 with the goal of studying backgrounds and place our first limits on spin-dependent interactions. 

\section{Head-tail effect in nuclear recoils from low-energy neutrons }
Studies of the nuclear recoils induced by low-energy neutrons from $^{252}$Cf have demonstrated that the DMTPC detector can measure the energy, direction, and versus of recoiling nuclei~\cite{Ref3}. Figure 1 (bottom left) shows a typical nuclear recoil reconstructed in the DMTPC detector at a pressure of 75 torr. The neutrons were 
traveling right to left.  
The decreasing dE/dx along the track direction is well visible, indicating the capability of determining the sense of the direction (``head-tail'') on an event-by-event basis. 
MC studies indicate excellent head-tail discrimination can be obtained for nuclear recoils above 70 keV in CF$_4$ at a pressure of  50 torr.

\section{Future detectors and expected sensitivity}

The DMTPC collaboration is designing a 1-m$^3$ detector to be operated in an underground site.  This detector has two 1-m$^2$ amplification planes. In the current design, each plane serves two 25-cm drift regions, and is imaged by 10 CCD cameras and 10 PMTs. By running this device for one year at 100 torr we will obtain an exposure of 100 kg-days. Assuming a threshold of 50 keV and passive neutron shielding, this device will allow us to set a limit on the spin-dependent cross section at $\sigma_p^{SD} \approx 10^{-38} cm^2$, as shown in 
figure 1 (right). 
This high sensitivity is achieved despite the limited mass due to the excellent sensitivity that fluorine has to spin-dependent interactions~\cite{Ref8} 
and because directionality increases the sensitivity to Dark Matter by over one order of magnitude~\cite{Ref9,Ref10}. 

\begin{figure*}[t]
\centering
\includegraphics[width=170mm]{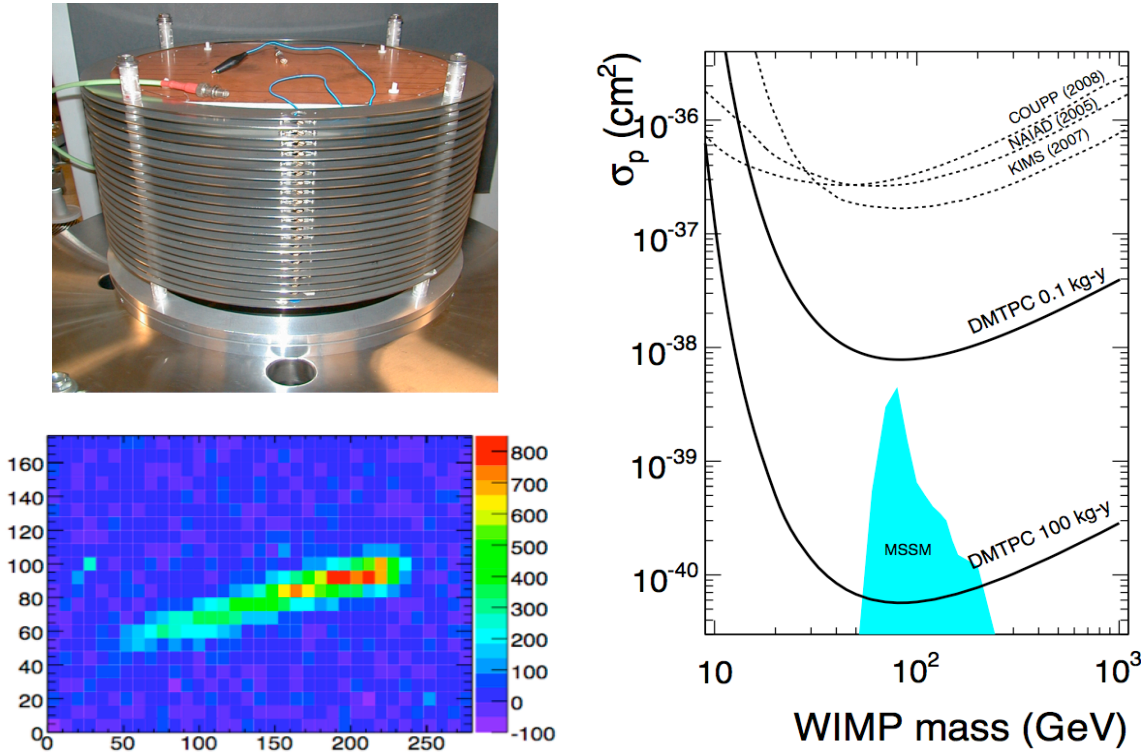}
\vspace{-0.5cm}
\caption{Top left: 10-liter DMTPC detector. The field cage and top mesh are visible. Bottom left: image of a nuclear recoil generated by a low-energy 
neutron traveling right to left. The larger dE/dx visible on the right of the recoil is consistent with
observation of the ``head-tail" effect. 
Right: limits on spin-dependent interactions of WIMPs on protons expected   for the 1-$m^3$ in a 1-year of underground data-taking 
as well as for 3 years operation of a larger DMTPC detector. 
} \label{fig1}
\end{figure*}

A larger detector with an active mass of a few hundred kg will explore a significant portion of the MSSM parameter space. The observation of directional WIMP signal by this detector will allow us to test our understanding of the local DM halo model. This detector is an ideal candidate for the DUSEL laboratory in South Dakota.

\begin{acknowledgments}
This work is supported by the Advanced Detector Research Program of the U.S. Department of Energy (contract number 6916448), the National Science Foundation, the Reed Award Program, the Ferry Fund, the Pappalardo Fellowship program, the MIT Kavli Institute for Astrophysics and Space Research, and the MIT Physics Department.
\end{acknowledgments}

\end{document}